\documentclass[doublecol]{epl2} 

\usepackage{graphicx}
\usepackage{amsmath, amssymb}
\graphicspath{Figs/}

\title{The role of interstitial gas in determining the impact response of granular beds} 

\author{John R. Royer\inst{1}\footnote{current address: Center for Soft Matter Research, New York University, New York, NY 10003.}, Bryan Conyers\inst{1}, Eric I. Corwin\inst{1}\footnote{current address: Department of Physics, University of Oregon, Eugene, OR 97403.}, Peter J. Eng\inst{1,2} \and Heinrich M. Jaeger\inst{1}}
\shortauthor{J. R. Royer \etal}

\institute{
\inst{1}James Franck Institute and Department of 
Physics, The University of Chicago, Chicago, IL 60637 \\
\inst{2}Consortium for Advanced Radiation Sources, The 
University of Chicago, Chicago, IL 60637}

\pacs{83.80.Fg}{Granular materials; rheology}
\pacs{45.70.Cc}{Compaction; granular systems}
\pacs{47.56.+r}{Fluid flow through porous media}

\abstract{We examine the impact of a solid sphere into a fine-grained granular bed.   Using high-speed X-ray radiography we track both the motion of the sphere and local changes in the bed packing fraction.   Varying the initial packing density as well as the ambient gas pressure,  we find a complete reversal in the effect of interstitial gas on the impact response of the bed:   The dynamic coupling between gas and grains allows for easier penetration in initially loose beds but impedes penetration in more densely packed beds.  High-speed imaging of the local packing density shows that these seemingly incongruous effects have a common origin in the resistance to bed packing changes caused by interstitial air. }

\begin{document}

\maketitle

\section{Introduction}

Granular materials often exhibit behavior intermediate between that of conventional solids and liquids.  Probing the resulting combination of liquid- and solid-like properties a number of recent studies investigated the impact of a large object into a bed of dry grains.    These studies focused on issues such as the drag on the impacting object \cite{Ciamarra:2004uq, Ambroso:2005uq, Ambroso:2005fk, Katsuragi:2007sp, Nelson:2008uq, Goldman:2008fq}, crater formation \cite{Uehara:2003bh,Bruyn:2004kx, Walsh:2003ul, Boudet:2006kx} the corona-like splash formed immediately after the impactor hits the bed surface \cite{Boudet:2006kx}, and the subsequent jet of grains formed by the collapse of the cavity left by the impactor \cite{Thoroddsen:2001lr, Lohse:2004qy, Royer:2005nx, Caballero:2007uq, Royer:2008fu, Kann:2010kx}.  So far however, almost all work considered the limit of loosely packed, marginally stable beds that readily compact in response to perturbations.   On the other hand,  densely packed beds must dilate in order for grains to move out of the way of inserted objects. This implies different resistance not only for slow, quasi-static perturbations  \cite{Schroter:2007cw} but also suggests that there should be a significant change in the dynamics for faster impacts.    

An important feature of the impact dynamics in granular systems is the coupling between the interstitial gas, typically air, and the grain packing.  For fine grained beds (grain diameters below $\sim$ 150 $\mu$m) this interaction can drastically change the impact dynamics.  In particular, in the presence of interstitial gas an impacting sphere penetrates much deeper than in vacuum  \cite{Royer:2007vn, Caballero:2007uq}.  One possible explanation, suggested by Caballero et al.\cite{Caballero:2007uq},  is drag reduction due to local fluidization provided by a layer of gas flowing around the moving sphere.   A different explanation relies on the presence of interstitial gas throughout the bed.  In situations where the gas permeability is low, i.e., fine-grained beds, the gas is effectively trapped during the short impact duration and can resist global packing density changes \cite{Royer:2008fu}.  For loose packings this leads to a response whereby a large portion of the bed surrounding the impact behaves similar to an incompressible liquid, allowing the impactor to sink in \cite{Royer:2007vn}.  On the other hand, for denser packings the same mechanism predicts that the presence of gas should impede the impactor movement, since the gas-particle coupling should tend to counteract dilation.  

Here we test for these two different scenarios by varying both the interstitial air pressure and the initial packing of the bed.  Using x-ray radiography, we track both the motion of the sphere through the bed and resulting local packing densities changes.

\section{Setup}

X-ray imaging was performed at the GeoSoilEnviroCARS  beam line at the Advanced Photon Source as in  refs. \cite{Royer:2007vn,Royer:2005nx,Royer:2008fu}.   The experimental setup and image processing are detailed in \cite{Royer:2008fu}, and we only outline key aspects here.   X-ray transmission through the bed was recorded at 6000 frames per second with resolution of 29 $\mu$m/pixel.  The beam size restricted the field of 
view to 22 mm x 8.7 mm sections of the bed, so in order to capture the dynamics across the full vertical extent of the bed, movies of multiple, independent sphere drops, were aligned and synchronized.  The detector was calibrated pixel by pixel to convert intensity to packing density, allowing us to correct  spatial variations in beam intensity and detector sensitivity.  

For each experiment, a steel sphere (diameter $D_s = $ 12mm) was dropped from a height of 35 cm into a bed of boron carbide 
($\textrm{B}_4\textrm{C}$) particles (diameter $d =$ 50 $\mu$m $\pm$ 10$\mu$m).   The bed was contained in a 
cylindrical polycarbonate tube with 35 mm inner diameter. Before 
each drop the bed was aerated by dry nitrogen entering through a diffuser built into the bottom of the container.   After
slowly turning off the nitrogen flow, the bed reproducibly settled to a loosely packed state with an initial average packing fraction $ \phi_0  =$ 0.51 $\pm 0.01$.   In this loose state the bed depth was 9 cm.   For experiments with denser beds, we compacted the bed by gently tapping the chamber walls until the top surface of the bed fell to fixed level.   The densest bed we obtained was 7.5 cm deep, corresponding to a 17\% decrease in volume.  With the x-rays we measured the initial packing of the dense bed to be $ \phi_0  =$ 0.60 $\pm 0.01$, in agreement with the change in height.    In both the loose and compacted beds, the initial packing varied by no more than 1-2\% across the bed height.   The system could be sealed and evacuated down to pressures of  0.7 kPa.  The pump speed was limited to prevent air from bubbling up and disturbing the packing.

\section{Results}

\begin{figure}
 \onefigure[width = 7.5cm]{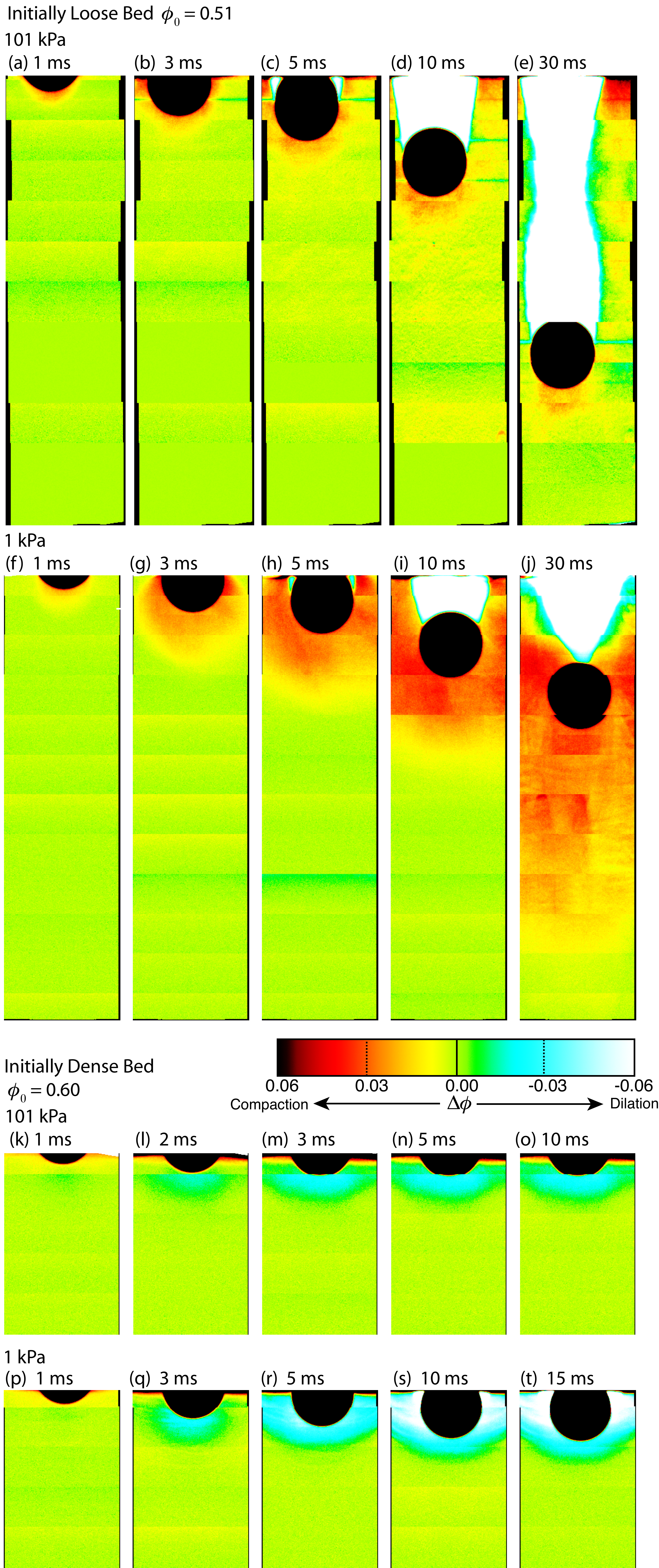}
\caption{ (Color) Composite x-ray images showing changes in local packing fraction $\Delta \phi = \phi -\phi_0$ after a metal sphere impacts a granular bed surface at time $t$ = 0.  The impact sequences (left to right) contrast initially loose beds, (a)-(e) and (f)-(j), with initially dense beds, (k) - (o) and (p) - (t).  For each the response under atmospheric pressure (101kPa) as well as vacuum (1kPa) are shown. The impacting sphere appears black while the cavity behind the sphere appears white.  Supplemental movies for each impact sequence are available online: Loose\_atmospheric.mov (Quicktime 2.1 MB), Loose\_vacuum.mov (Quicktime 2.3 MB), Dense\_atmospheric.mov (Quicktime 852KB) and Dense\_vacuum.mov (Quicktime 373KB).}  

\label{fig:xray}
\end{figure} 

\begin{figure}
 \onefigure[width = 7.5cm]{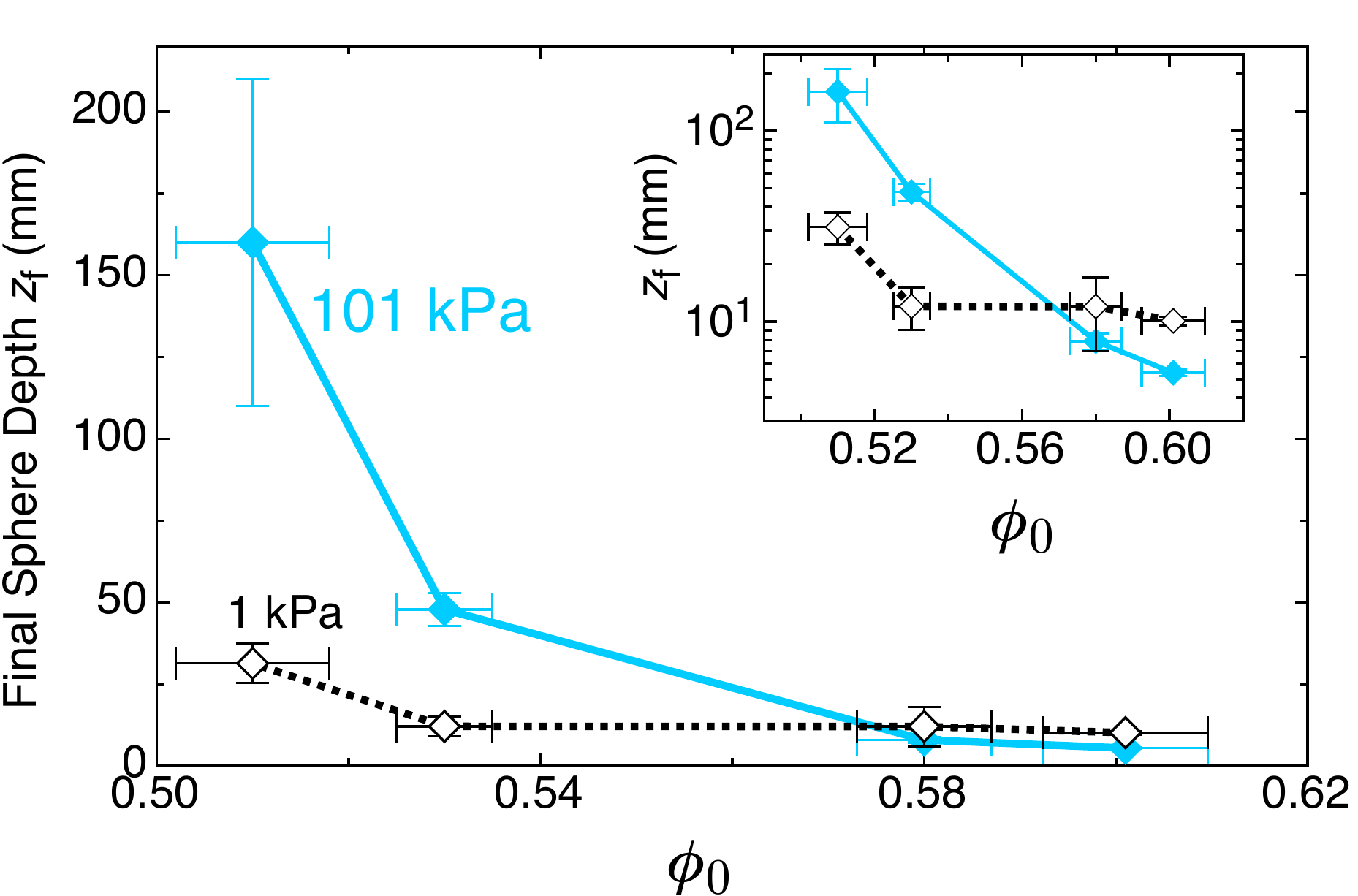}
\caption{ Transition in impact dynamics.   The final penetration depth of the sphere, $z_{f}$, as a function of initial packing fraction $\phi_0$ at $P_0 = 101$ kPa (solid symbols) and in vacuum $P_0 = 0.9$ kPa (open symbols).    For the loose ($\phi_0 = 0.51$) bed at 101 kPa, the sphere hit the bottom of the bed in the chamber used for x-ray measurements so $z_{sf}$ was measured separately in a deeper bed using a line attached to the end of the sphere.  Inset:  The same data on a log scale to highlight the crossing around $\phi_0$ = 0.58.  }  

\label{fig:transition}
\end{figure} 

\begin{figure}
 \onefigure[width = 7.5cm]{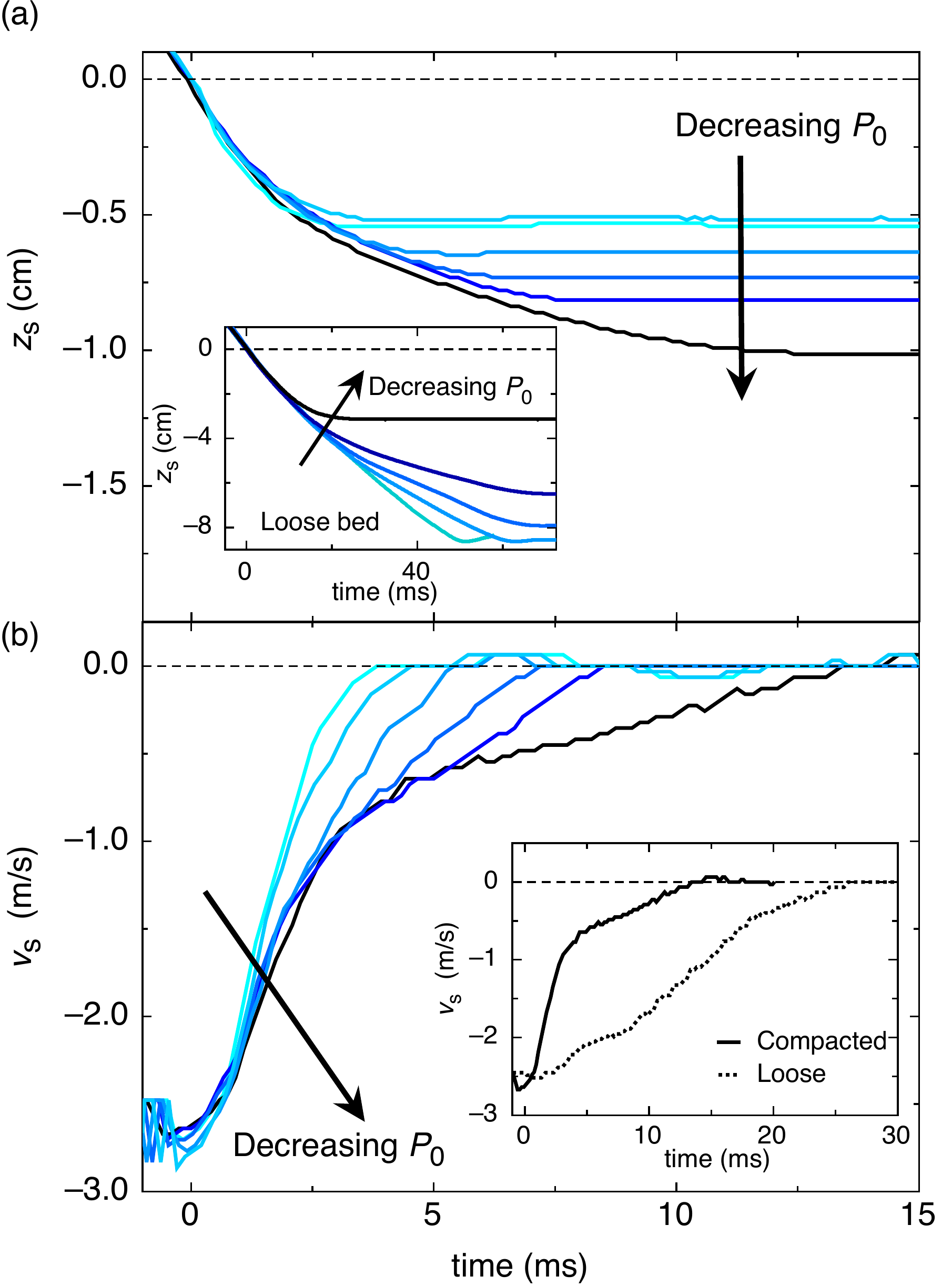}
\caption{ Sphere dynamics.    (Color Online).  (a) Vertical position $z_s(t)$ of the bottom tip of the sphere as a function of time after impact ($t = 0$ s) for an initially dense bed $\phi_0$ = 0.61.   From top to bottom: $P_0 = $ 101 kPa, 50.6 kPa, 14.3 kPa, 6.7 kPa, 3.6 kPa and 0.7 kPa.  Inset:  $z_s(t)$ for an initially loose bed $\phi_0$ = 0.51 and from bottom to top: $P_0 = $ 101 kPa, 12 kPa, 8.7 kPa, 4.9 kPa and 0.7 kPa.    (b)  Velocity $v_s(t)$ computed from curves in (a) for $\phi_0 = 0.61$.  Inset:  Comparison of $v_s(t)$ in vacuum ($P_0 < 1$ kPa) for initially loose and compacted beds.   }
\label{fig:ball_track}

\end{figure} 

\begin{figure}
 \onefigure[width = 8.5cm]{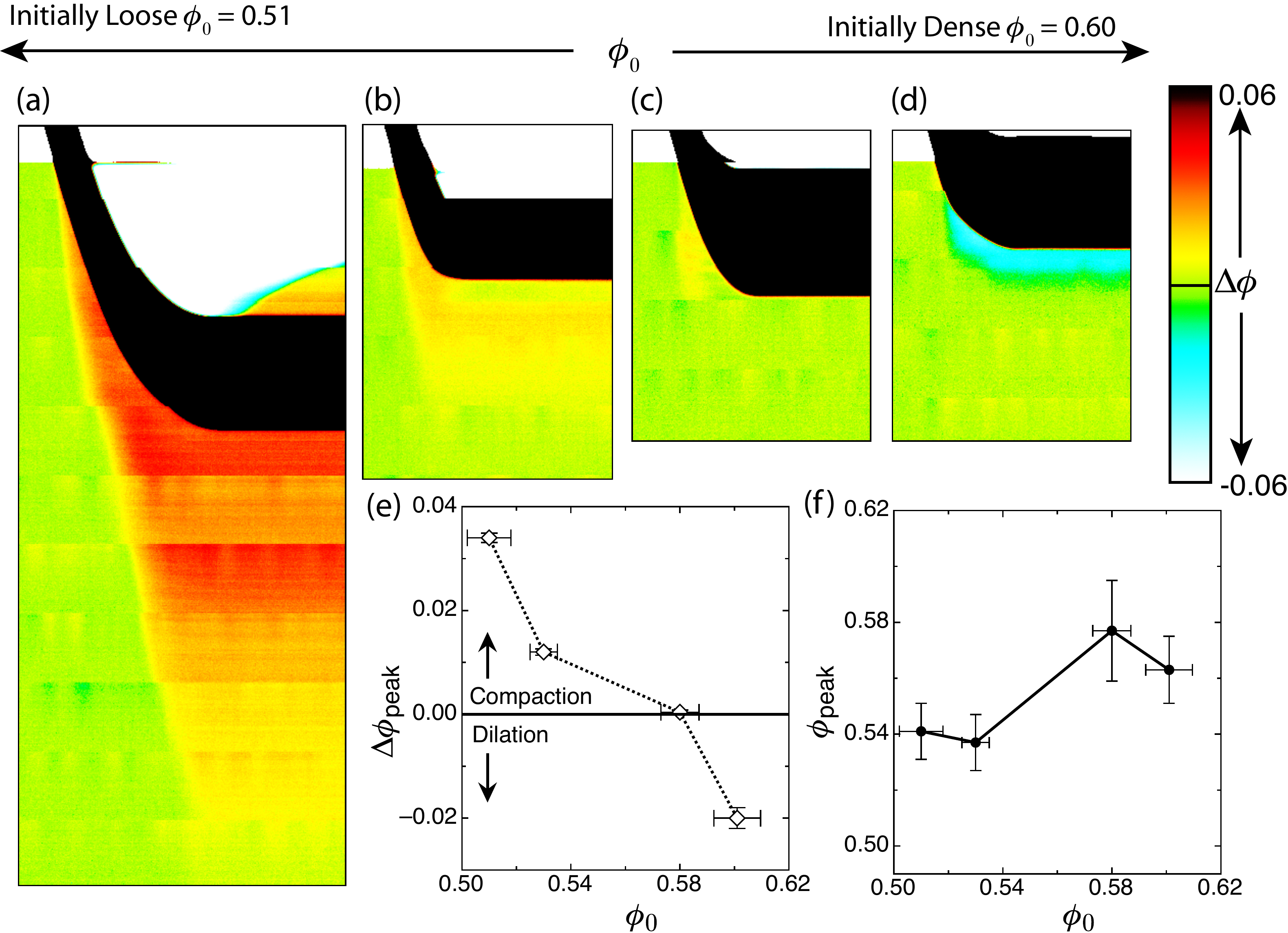}
\caption{ Packing changes in vacuum.    (a-d) Space-time plots of the centerline of the x-ray movies in Fig. 1, showing the change in packing $\Delta \phi$ in vacuum ($P_0 < 1$ kPa) for different initial packing densities (a) $\phi_0 =$ 0.51 $\pm$ 0.01(b) $\phi_0 =$ 0.53 $\pm$ 0.01(c) $\phi_0 =$ 0.58 $\pm$ 0.02, (d) $\phi_0 =$ 0.60 $\pm$ 0.01.  (e-f)  Packing density observed  below the sphere after it has come to rest:  peak change (e) and peak value (f) for different initial packing fractions}  
\label{fig:front}
\end{figure}

In Fig. \ref{fig:xray} we show the motion of the sphere and the local change in bed packing density $\Delta \phi = \phi(t) - \phi_0$ during penetration of initially loose and initially dense beds at atmospheric pressure ($P_0$ = 101 kPa) and in vacuum ($P_0 <$ 1 kPa).  These images reveal striking differences in the impact dynamics. 

In the initially loose bed under atmospheric pressure (Figs. \ref{fig:xray} (a) - (e)) the sphere easily sinks into the bed, leaving a large cavity behind it.  In fact, the sphere reaches the bottom of the 9 cm deep bed and even bounces before coming to rest.   In the loose bed under vacuum (Fig. \ref{fig:xray}f-j) the sphere still penetrates into the bed and opens up a cavity, but the resistance of the bed is substantially increased.   In contrast to the loose bed under atmospheric pressure, the sphere only sinks 3.5 cm below the top surface and the resultant cavity is much smaller.    

In the initially dense bed the resistance to penetration is significantly increased, stopping the sphere before it can even sink below the top surface.   However, now the effect of the interstitial gas is reversed: in vacuum the sphere sinks nearly twice as deep  (Fig. \ref{fig:xray}p-t) as in atmospheric pressure (Fig. \ref{fig:xray}k-o).  To quantify this pressure dependence,  we track the sphere position $z_s(t)$ varying the ambient pressure (Fig. \ref{fig:ball_track} a).   These trajectories show that the penetration depth monotonically increases with decreasing ambient pressure $P_0$ in dense beds.   In loose beds, by contrast, the penetration depth increases with increasing ambient pressure (Fig. \ref{fig:ball_track} a inset) \cite{Royer:2007vn, Caballero:2007uq}.

To further examine this reversal in the impact dynamics, we plot the final sphere depth $z_f$ against $\phi_0$ for beds at atmospheric pressure and under vacuum in Fig. \ref{fig:transition}.    At atmospheric pressure, the final sphere depth depends strongly on the initial packing density, decreasing by over an order of magnitude as this density is increased from $\phi_0$ = 0.51 to 0.62.  Under vacuum, on the other hand, the penetration depth is nearly independent of $\phi_0$, with only a slight increase at our lowest value $\phi_0$ = 0.51.   The inset in Fig. \ref{fig:transition} shows the same data on a log-log scale to highlight the crossover from deeper penetration under atmospheric pressure to deeper penetration under vacuum.  Though our bed preparation method is not optimal for controlling the packing fraction precisely and we only have results for four values of $\phi_0$, the data indicate that the crossover occurs near $\phi_0$ = 0.57-0.58.

In the initially loose bed, during impact under vacuum a large front of compacted grains builds up ahead the sphere (Fig. \ref{fig:xray}f-j).   The extent of the compacted region grows as the sphere moves through the bed, so the edge of this compaction front travels faster than the sphere.    However, after the sphere comes to rest the front does not continue to propagate and instead stops before reaching the bottom of the bed (Fig. \ref{fig:xray} j).    This differentiates this compaction front from propagating pulses observed, e.g., in granular chains \cite{Sen:1998yb}.   The maximum packing density in the compaction front is $\phi \sim$ 0.54, much less than the packing densities of up to $\phi \sim 0.61$ obtained by manually tapping the chamber to compact the bed.  

In the initially dense bed under vacuum, there is considerable dilation around the moving sphere, as one would expect  \cite{Reynolds:1885mj, Onoda:1990yk}.   Just like the compaction front in the loose bed, the dilation front does not propagate through the bed after the sphere has come to rest.   The packing fraction just below the sphere in the dilated region is about $\phi \sim $ 0.56, just slightly higher than the 0.54 measured for the compaction front in the loose bed.  

For both loose and dense beds under atmospheric pressure, most of the changes in packing density occur in a small region just ahead of the sphere (Figs. \ref{fig:xray}a-e, k-o).   This is in line with the idea that  the interstitial air  resists changes in the packing density. An initially loose, marginally stable bed thus stays liquid-like and flows out of the way of the sphere \cite{Royer:2007vn}, while an initially dense bed stays solid-like and flows very little.  

To quantify how the resistance of the dense bed to penetration depends on the ambient pressure, we  compute $v_s  = dz_s/dt$ from the measured trajectories $z_s(t)$, shown in Fig. \ref{fig:ball_track} b.    There is a sharp, initial deceleration which is nearly independent of the ambient pressure.   At reduced ambient pressure, at later times there is turnover to a lower deceleration.    In the inset in Fig. \ref{fig:ball_track} b we compare the sphere velocity in vacuum in an initially loose and initially compacted bed.   After the initial rapid deceleration in the dense bed, the deceleration at later times is very close to the deceleration in the loose bed.      

The connection between packing density changes and impactor motion is most clearly seen in space time plots from composite x-ray movies under vacuum at varying initial packing fractions (Fig.  \ref{fig:front}a-d).   In the loosest bed   Fig.  \ref{fig:front}a there is a large, clear compaction front which extends far ahead of the sphere, but this front dies out and stops at roughly the same time the sphere comes to rest.   As the initial packing is increased  (Fig.  \ref{fig:front}b), the sphere does not penetrate as deep and the compaction front becomes less pronounced until around $\phi_0=$ 0.58 there is almost no local density change as the sphere move through the bed  (Fig. \ref{fig:front}c).   Finally, in the densest bed ( Fig.  \ref{fig:front}d) we see that the dilation front ahead of the sphere stops the same time the sphere comes to rest.   Plotting the peak change in packing below the sphere (Fig.  \ref{fig:front}e) we can track this change from compaction in loose beds to dilation in dense beds.   At the same time, the magnitude of the packing fraction at this peak does not exhibit any clear dependence on $\phi_0$, varying between 0.54 and 0.58  (Fig. \ref{fig:front}f).   

\section{Discussion and Conclusions}

Our results demonstrate that, during sudden impact, the primary role of the interstitial air in a fine-grained granular bed is to oppose changes in the bed packing density.   In loose beds the trapped gas prevents compaction of the grain arrangement, and as a result the bed flows out of the way of the intruder, similar to an incompressible fluid.  This allows for relatively easy penetration and explain why a heavy sphere, even if started just above the bed with zero impact velocity, can sink in deeply \cite{Lohse:2004qy}.   In sufficiently dense beds, the trapped gas has the reverse effect, working against dilation of the grain arrangement and thereby resisting penetration.  

In the absence of interstitial gas, the impacting object can change the grain packing density much more effectively, generating a pronounced compaction front if the bed is loose and dilating the bed when it is dense.  For loose beds, in particular, the result is behavior similar to that observed ahead of a plow pushed into soil or snow, as seen by the reddish-colored regions of enhanced density in Figs. 1f-j and 4a. 

If local, gas-mediated fluidization around the impacting sphere were to provide the mechanism for deep penetration in loosely packed beds, then the presence of gas should have allowed for deeper impact also in more densely packed beds.  The fact that we find the opposite behavior, whereby gas reduces the impact depth in the densely packed case, appears to rule out this scenario.  Indeed,  in qualitative terms the same reversal in relative penetration depth   with increasing packing density was already pointed out  in a NASA Technical Note in 1963 \cite{Clark:1963uq}

As the packing density is increased, we observe a gradual transition  from compacting to dilating response of the bed.  For the type of  bed material examined here, the crossover  between these two regimes  occurs at a packing fraction $\phi \sim$ 0.58  (Fig. 4e).    Within our experimental accuracy, this agrees well with the packing fraction $\phi \sim$ 0.57-0.58 at which the penetration depth with ambient air present becomes smaller than the same depth measured in vacuum (Fig. 2 inset).  This finding says that the crossover from compaction to dilation occurs at, or at least very close to, the density at which the role of interstitial gas is negligible.  It suggests, therefore, that the particular crossover value should be independent of gas-grain interactions.  

In principle, this makes comparison possible with experiments or calculations where the coupling to air does not need to be considered, such as slow, quasi-static penetration.  On general grounds we expect the crossover density to lie above the value $\phi_{rlp}$ for random loose packing, since $\phi_{rlp}$ corresponds to marginally stable configurations that collapse and compact when perturbed \cite{Jerkins:2008vn}.  For slowly settled non-cohesive spheres, $\phi_{rcp}\sim$0.56-0.57, with lower values for rougher particles and larger values for larger mismatch with the suspending fluid \cite{Onoda:1990yk, Jerkins:2008vn}.  A few percent above  $\phi_{rcp}$  Schr\"{o}ter et al.  indeed observed a transition in the resistance of a granular bed to quasi-static penetration by a rod  \cite{Schroter:2007cw}. Using spherical glass beads Schr\"{o}ter at al. found this transition to occur  at values just below  0.60.  In our experiments, the deviation from sphericity in the boron carbide particles is likely to be responsible for pushing  $\phi_{rlp}$ to lower values and with it the crossover density to around 0.58.

We thank  M. Rivers, M. M\"{o}bius and S. Nagel for insightful discussions. This work was supported by the W. M. Keck Foundation through the Keck Initiative for Ultrafast Imaging at Chicago and by the NSF through its MRSEC program under DMR-02137452 and DMR-0820054.. GSECARS is supported by NSF EAR-0217473, DOE DE-FG02-94ER14466, the State of Illinois and the W. M. Keck Foundation. Use of the Advanced Photon Source was supported by DOE-BES under W-31-109-Eng-38.

\bibliography{ref_decom.bib}
\bibliographystyle{eplbib}

\end{document}